\documentclass[twocolumn]{jpsj2} 
\title{Contact and Quasi-Static Impact of a Dissipationless Mechanical Model}

\author{Hiroto \textsc{Kuninaka}$^{1}$\thanks{E-mail address:
kuninaka@kuchem.kyoto-u.ac.jp} and Hisao \textsc{Hayakawa}$^{2}$
}

\inst{$^{1}$Department of Chemistry, Kyoto University, Sakyo-ku, Kyoto, 606-8502 \\
$^{2}$Department of Physics, Yoshida-south campus, Kyoto University, Sakyo-ku, Kyoto, 606-8501 }

\abst{Collisions and contacts of elastic materials are numerically and 
theoretically investigated. Using a two-dimensional spring-mass model 
with defect particles under the free boundary condition, 
we reproduce the Hertzian contact theory at equilibrium 
and the quasi-static theory for low speed impacts.}

\kword{Contact problems, Collisions, Quasi-static theory, Hertzian contact theory, 
The second law of thermodynamics}

\begin{document}
\maketitle
\newpage
\section{INTRODUCTION}
The origin of irreversibility from a reversible mechanical model 
is one of the most fundamental subjects in non-equilibrium statistical 
mechanics. A typical example can be seen in transport phenomena in nonlinear 
spring models\cite{lepri,casati}. Although most of models discuss the 
transport processes under the influence of thermostats, we still 
do not understand the mechanism to reach an equilibrium state based on 
a purely mechanical model. 

The aim of this paper is to derive macroscopic laws 
of elastic materials based on a microscopic lattice model. 
Here we discuss contacts and impacts between 
elastic materials. 
The former is related to the origin of the second law of thermodynamics 
from a purely mechanical model. 

For contacts of elastic materials, 
we believe that the contacts between elastic bodies can be described by 
the Hertzian contact theory\cite{hertz,landau,love}. 
The two-dimensional Hertzian contact theory gives us the relation between 
the deformation of a disk $\delta$ and the compressive force $P$ as
\begin{equation}\label{hertz_cnt}
\delta \simeq \frac{P}{\pi E^{*}}
\left\{ \ln \left( \frac{4\pi E^{*} R}{P}\right)-1-\nu\right\},
\end{equation}
where $R$ is the radius of the undeformed disk\cite{gerl}. 
Here, $E^{*}$ is 
the reduced Young's modulus, $E^{*}=E/(1-\nu^{2})$, where  
$E$ and $\nu$ are Young's modulus and Poisson's ratio, respectively. 

For the low speed head-on collisions of elastic materials, 
the relation between the restitution coefficient $e$ 
and the impact speed $v$ is described 
by the quasi-static theory of low speed impacts\cite{kuwabara,brilliantov96}. 
The quasi-static theory is an extension of 
the Hertzian contact theory to include the internal viscosity of materials. 
By solving the equation of motion for the deformation 
with adequate initial conditions and calculating the rebound speed, 
we can obtain the relation between the impact speed $v$ and 
the restitution coefficient $e$. 


The restitution coefficient $e$ depends also 
on the incident angle\cite{louge,sundararajan}. 
We have recently carried out a two-dimensional simulation of oblique impacts 
using a dissipationless elastic model based on the mass-spring model
\cite{kuninaka_prl,phase_trans} 
to reproduce and explain the previous experimental result 
by Louge and Adams\cite{louge}. 
Although the model can reproduce similar results to the experimental 
results, one of the apparent defects in the model is 
that the system cannot reach an equilibrium state 
described by the Hertzian contact theory\cite{hertz,landau,love}. 
When we put the disk on the wall under the influence of the gravity, 
the release of initial potential energy is recurrent 
and the system keeps oscillations. 
We also have expected that $e$ can be described by the quasi-static theory 
for low speed head-on collision of the disk with a potential wall, 
but we have found a discrepancy between the numerical result and 
the prediction by the quasi-static theory\cite{gerl,ces}. 
Refining the model to overcome those defects in a dissipationless model 
will be helpful to understanding the origin of 
irreversibility and the second law of thermodynamics 
from a mechanical point of view.

In this paper, 
we propose a microscopic dissipationless model to reproduce 
the Hertzian contact at equilibrium and the quasi-static theory 
for low-speed impacts. These results enable us to understand 
the origin of irreversibility for finite degrees of freedom  
from a reversible mechanical model.  
The construction of this paper is as follows. In the next section we introduce 
our model. In Sec. 3 and 4 we show the results of our simulation of 
contact problems and low-speed impacts, respectively, and briefly summarize 
our results. Appendix A is devoted to the calculation of elastic moduli 
of our model. 
\begin{figure}[htbp]
  \begin{center}
    \includegraphics[width=.4\textwidth]{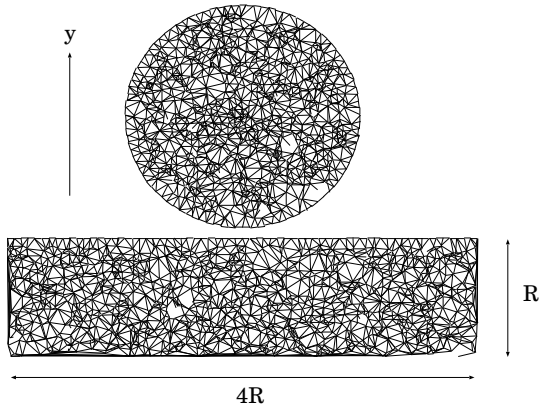}
  \end{center}
  \caption{The elastic disk and wall consisted of random lattice. 
    We set  $y=0$ at the surface of the wall. 
    In this figure, the height and the width of the wall 
    are respectively $R$ and $4R$ with the undeformed radius of the disk $R$.}
  \label{fig2-1}
\end{figure}

\section{Model}

Let us introduce our model(Fig.\ref{fig2-1}).
It is basically the same as our previous model 
which obeys Hamilton equation\cite{kuninaka_jpsj2003,kuninaka_prl}. 
The disk consists of $1099$ mass points while the wall consists of $1269$ 
mass points. 
The two corner points of the bottom of the wall are fixed. 
The surfaces of both the disk and the wall are initially flat. 
The interaction between the disk and the wall is introduced as follows. 
Each mass point $i$ on the lower half boundary of the disk receives the force, 
${\bf F}(l_{s}^{(i)})=aV_0\exp(-a l_{s}^{(i)}){\bf n}_{s}^{(i)}$, 
where $l_{s}^{(i)}$ is the distance between $i$-th surface mass point 
of the disk and the nearest surface spring of the wall, $a=500/R$, 
$V_0=amc^2R/2$, $m$ is the mass of each mass point $i$, 
$c$ is the one-dimensional speed of sound, 
and ${\bf n}_{s}^{(i)}$ is the unit vector normal to the connection 
between 
two surface mass points of the wall\cite{kuninaka_jpsj2003,kuninaka_prl}. 
Thus, the dynamical equation of motion for each mass point $i$ 
of the lower half boundary of the disk is described by 
\begin{equation}\label{keq}
m \frac{d^{2}{\bf r}_{i}}{d t^{2}}
=\sum^{N_{i}}_{j=1} 
\left\{-k_{a} {\bf x}_{ij} - k_{b} {\bf x}_{ij}^{3}\right\} 
+\Theta(l_{th}-l_{s}^{(i)}) a V_0 \exp(-a l_{s}^{(i)}){\bf n}_{s}^{(i)},
\end{equation}
where ${\bf r}_{i}$ is the position of $i$-th mass point, 
$t$ is the time, 
$N_{i}$ is the number of mass points connected to $i$-th mass point, 
${\bf x}_{ij}$ is the relative deformation vector of the spring 
from the natural length between $i$-th and $j$-th connected mass points, 
$k_{a}$ and $k_{b}=k_{a}\times10^{-3}/R^{2}$ are the spring constants. 
Here $\Theta(x)$ is the step function, i.e. $\Theta(x)=1$ 
for $x \ge 0$ and $\Theta(x)=0$ for $x < 0$, 
and the threshold length $l_{th}$ is the average of 
the natural lengths of the springs of the disk. 
For internal mass points, the last term of the right hand side 
of eq.(\ref{keq}) is omitted. In most of our simulations, 
we adopt $k_{a} = k^{(d)}_{a} =1.0 \times m c^2/R^2$ 
for the disk and $k_{a} = k^{(w)}_{a} =1.0 \times 10^{2} m c^2/R^2$ 
for the wall. 
Numerical integration of eq.(\ref{keq}) is performed 
by the fourth order symplectic integrator with the time step $dt=10^{-3}R/c$.

There are two main differences between our new model and the previous model. 
Here we introduce some defect particles and adopt 
the free boundary condition in which the kinetic energy 
can escape from the boundary. 
The defect particles are introduced by choosing mass point $i$ 
at random and eliminating $N_{i}-1$ connecting bonds 
among $N_i$ connecting bonds to $i$-th mass point. 
We have introduced $10$ defects for each body. 
The reason to choose $10$ defects will be discussed later.  
With the introduction of these defects, 
we expect that the motion of defects becomes irregular and 
the vibrating wave can be localized without spreading. 
This irregular motion may create the irreversibility of 
time evolution of mass points. 
In addition, we adopt the free boundary condition 
for both sides and the bottom of the wall in contrast to 
the reflective boundary condition in our previous model. 
We regard the wall as a part of a large system. When we put a 
mechanical perturbation in the wall such as a contact or an impact, 
the effect propagates as elastic waves and goes out from the boundary 
because our system is a part of a larger system. 
In this situation, ${\bf J} \cdot {\bf n}_{b} > 0$ should be satisfied where 
${\bf n}_b$ and ${\bf J}$ are respectively the unit normal vector on 
the boundary and the energy flux 
${\bf J}=\sum_{i}\{(1/2)m{\bf v}_{i}^{2}+\epsilon_{i}\}{\bf v}_{i}$ 
with the potential energy $\epsilon_{i}$. 
In order to realize the condition, 
at each time step of numerical integration, 
we set ${\bf v}_{i}={\bf 0}$ 
if ${\bf v}_{i}$ is directed to inner region of the wall. 
Because the reflected wave does not come into the system,   
the total energy of our system is not conserved. 

Here we briefly comment on the boundary condition we adopt. 
Our model does not conserve the total energy because there is 
the outgoing energy flux from the system. 
We can put thermostats, e.g. Nos{\'e}-Hoover thermostats\cite{nose}, 
on the boundary to keep the energy conservation law. However, 
once we introduce the thermostat in the system, the phase volume is 
not conserved, and there is the entropy production\cite{evans}. 
In addition, we believe that the thermalization from the thermostat 
does not play an important role in macroscopic elastic materials\cite{ces}. 
Thus, we adopt the model without the conservation of the energy. 
In the next section we will show that the introduction of the thermostat 
does not make much difference in our results. 
\section{Simulation of Contact Problems}

In this section, we show the results of our simulation 
of elastic contacts. 
At first, we show the results of the simulation for the contact problem. 
\begin{figure}[htbp]
  \begin{center}
    \includegraphics[width=.45\textwidth]{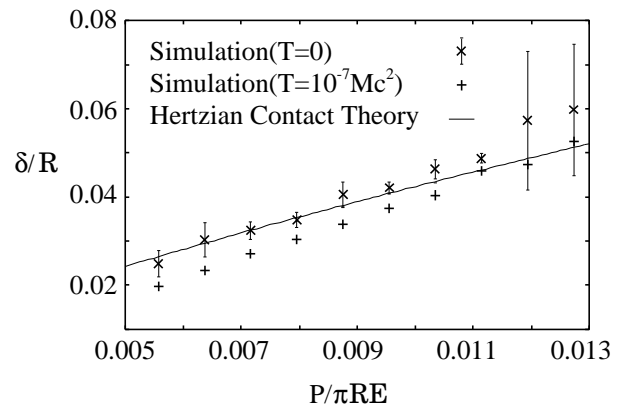}
  \end{center}
  \caption{The relation between the deformation and the external force. 
    Cross points with error bars are the numerical results with 
    the initial temperature $T=0$. Plus points are the numerical results with 
    the initial temperature $T=10^{-7}Mc^{2}$ in the simulation with 
    Nos{\'e}-Hoover thermostats. 
    The solid line is the prediction by the Hertzian contact theory 
    with $\nu=0.336$.}
  \label{htz}
\end{figure}
In this simulation, we introduce the wall whose height and width are 
$4R$ and $R$, respectively. 
We put the disk in the external field $P$ ranging 
from $P=5.77 \times 10^{-3}\pi R E^{*}$ to $P=1.27 \times 10^{-2}\pi R E^{*}$. 
Thus, in this case, we add the term $-(P/N) {\hat {\bf y}}$  
in the right hand side of eq.(\ref{keq}), 
where $N=1099$ and ${\hat {\bf y}}$ is the unit vector in $y$ direction. 
The initial oscillation of the center of mass of the disk relaxes 
to reach a stable oscillation. After the relaxation of the center of mass, 
we calculate the deformation of the disk, 
$\delta \equiv R-R_d$, where $R_d$ is the deformed radius of the disk 
which is the distance from the contact patch to 
the center of mass of the disk.

Figure \ref{htz} is the relation between $\delta/R$ and $P/\pi R E^{*}$. 
Cross points are the averaged results of $10$ disks 
with different configuration of mass points, 
and the error bars are the standard deviations.
The solid line in Fig.\ref{htz} is eq.(\ref{hertz_cnt}) with $\nu=0.336$ 
calculated from the two-dimensional theory of elasticity 
based on the assumption of an isotropic disk. Details of the calculation of 
elastic constants of the model are shown in Appendix A\ref{appA}.
Our simulation data are well reproduced by the theoretical curve 
which does not include any fitting parameters. 
Thus, we conclude that our model can 
produce an equilibrium state at the contact.  

Let us discuss the effect of thermalization from 
thermostats on our results. We have carried out another simulation, in which 
the wall has fixed boundaries at the bottom and the both side ends, and 
the mass points connected to the fixed boundaries obey the equation of motion 
of particles connected to the Nos{\'e}-Hoover thermostat\cite{nose}.  
We have arranged that the temperature of the wall becomes 
$T=10^{-7} M c^{2}$.
Here we define the temperature of the thermostat by the variance of 
the Maxwell-Boltzmann distribution which we give as the initial velocity 
distribution of the thermostat particles. 
Plus points in Fig. \ref{htz} are 
the averaged results of $10$ disks with different configuration of mass points 
in the simulation with the Nos{\' e}-Hoover thermostat. 
This result shows that 
the introduction of thermostats does not affect the relation between 
the compressive force and the deformation so much. 
From this result, we conclude that 
the thermlization from the thermostat does not play an important role in our 
simulation of contact problems. 

Introducing defect particles in the model plays an important role in 
the relaxation of internal vibration in the contact problem. 
To characterize the relaxation process to an equilibrium state, 
we investigated the time evolution of the velocity distribution function (VDF)
$f(v_x,v_y,t)$, where $v_x$ and $v_y$ are velocity components of 
mass points of the disk, and the Shannon entropy when 
the compressive force $P=8.75 \times 10^{-3}\pi R E^{*}$.  
To obtain the velocity distribution function 
$f(v_x,v_y,t) \equiv \sum_{x}\sum_{y}f(x,y,v_x,v_y,t)$, 
we calculate $f(x,y,v_x,v_y,t)$ by averaging 
the data from $t-6 R/c$ to $t$.  
\begin{figure}[htbp]
  \begin{center}
    \includegraphics[width=.4\textwidth]{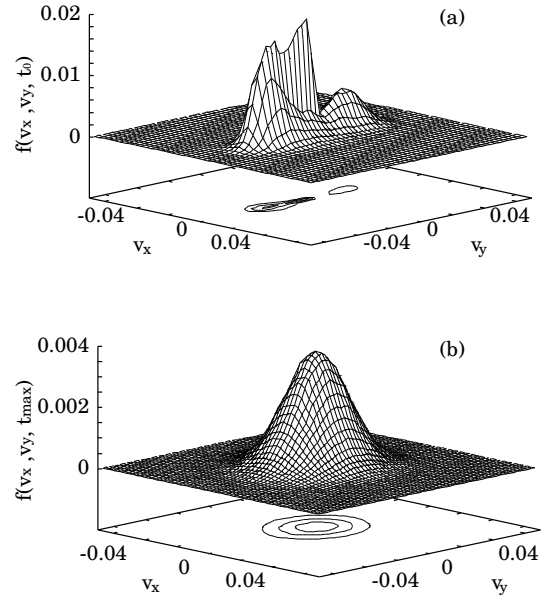}
  \end{center}
  \caption{The velocity distribution of mass points of the disk at  
(a) $t=t_{0}$ and (b) $t=t_{max}$.}
  \label{binv9bw}
\end{figure}
\begin{figure}[htbp]
  \begin{center}
    \includegraphics[width=.4\textwidth]{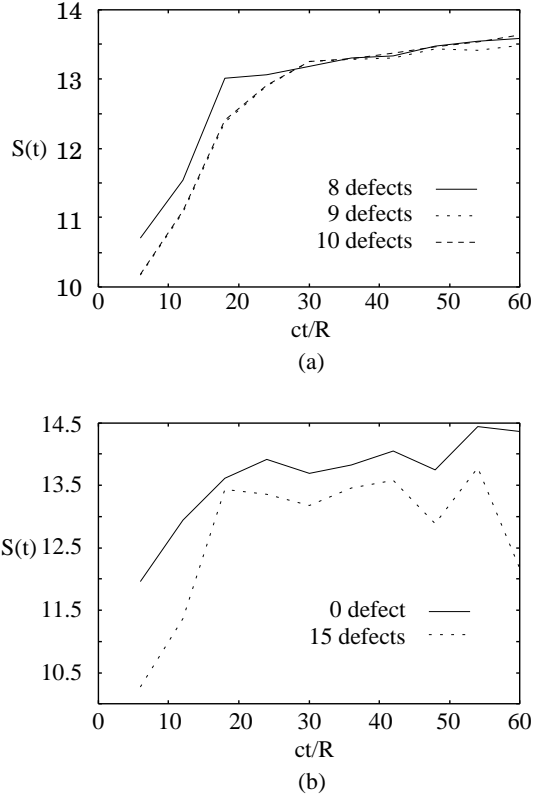}
  \end{center}
  \caption{The time evolutions of the Shannon entropy $S(t)$ when the 
    number of defects are (a) from 8 to 10 and (b) 0 and 15. 
    X-axis is the scaled time while y-axis 
    is the Shannon entropy $S(t)$ calculated from the distribution function 
    of internal mass points of the disk.}
  \label{ent}
\end{figure}

Figure \ref{binv9bw} shows the time evolution of VDF $f(v_x,v_y,t)$. 
The distribution function near the initial stage (Fig. \ref{binv9bw}(a)), 
$f(v_x,v_y,t_{0}=6 R/c)$, 
has three peaks along the axis $v_x=0$. 
The peaks may be attributed to the coexistence of the reflective motion  
and the compressive motion of mass points in the disk. 
Figure \ref{binv9bw}(b) shows that $f(v_x,v_y,t_{max}=60 R/c)$ 
has a Gaussian form. 
In fact, 
$f(v_x,0,t_{max})$ can be well fitted by the Gaussian whose variance is $0.02$. 

In addition, 
we have investigated the time evolution of the Shannon entropy. 
Figure \ref{ent} shows 
the time evolution of the entropy defined by 
$S(t) \equiv -\sum_{x}\sum_{y}\sum_{v_x}\sum_{v_y}f(x,y,v_x,v_y,t) 
\ln f(x,y,v_x,v_y,t)$ 
for several numbers of defects. 
We have investigated the dependence of 
the number of defect particles on the entropy. 
In Fig.\ref{ent}, 
we change the number of defects in the disk from $0$ to $15$, 
and investigate the time evolution of the entropy for each disk. 
In the case of the disks whose number of defects are from $8$ to $10$, 
after $t=30 R/c$, 
the time evolution of the entropy is described as an universal curve 
which shows monotonic increase and reaches the maximum value $13.5$. 
In the case of other disks, 
the entropy does not show the monotonic increase. 
The mixing effect is not enough for the smaller number of defects 
while a global oscillation is excited because of the weakness of the disk 
structure for too large number of defects.

\begin{figure}[htbp]
  \begin{center}
    \includegraphics[width=.45\textwidth]{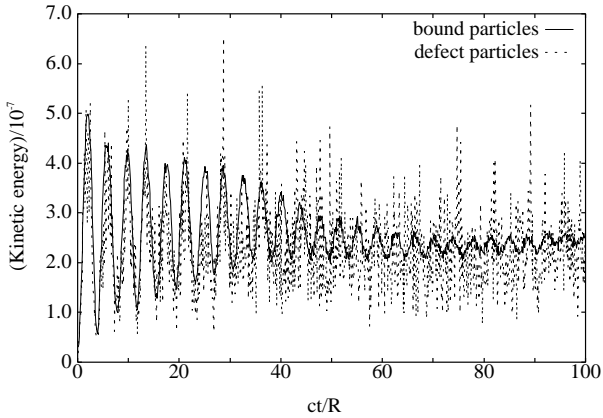}
  \end{center}
  \caption{The time evolution of the kinetic energy per one particle 
    of bound particles and defect particles of the disk.}
  \label{ene}
\end{figure}

We have also calculated the averaged kinetic energy per one defect particle 
(Fig.\ref{ene}). 
The time averaged kinetic energy of one defect is not so large as that of 
the other particles. However, the kinetic energy for one defect  
changes at random with larger amplitude. 
This tendency originates from irregular motion of the defect particles,  
which enables the oscillation modes to mix and to reach an equilibrium state. 
In realistic systems, such a phenomenon can be seen in phonon scattering by 
defects and impurities of solids, the scattering which realizes 
a thermal equilibrium state. 
We conclude that the irregular motion of the defect particles generates 
irreversibility and makes the model reach an equilibrium state, 
which reproduces the Hertzian contact of our model. 

\section{Simulation of Low-speed Impact}

In this section, we show the results of the simulation of low speed impact. 
Here we arrange that the height and the width of the wall are respectively 
$R$ and $4R$. 
From the simulation of the head-on collision of the disk with the wall, 
we calculate $e$ for each initial speed. 
The initial speed is ranged from $v=1.0 \times 10^{-3} c$ 
to $v=1.0 \times 10^{-2} c$ without the external field.

\begin{figure}[htbp]
  \begin{center}
    \includegraphics[width=.45\textwidth]{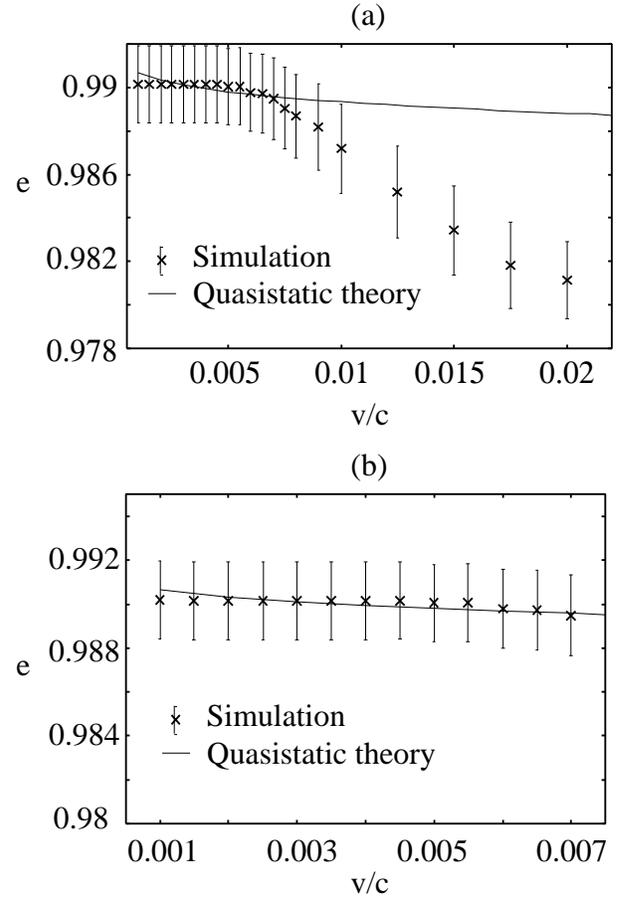}
  \end{center}
  \caption{The relation between the colliding speed $v$ 
    and the restitution coefficient $e$. 
    The solid line indicates the prediction by the quasi-static theory. 
    The cross points with error bars are numerical results in the range 
    (a) from $v/c=0.001$ to $v/c=0.02$ and (b) from $v/c=0.001$  to $v/c=0.007$.}
  \label{quasi}
\end{figure}

Figure \ref{quasi} (a) is the relation between the restitution 
coefficient $e$ and the impact speed $v/c$ ranging from $v/c=0.001$ to 
$v/c=0.02$ while Fig. \ref{quasi} (b) is the results in the low colliding speed. 
The cross points  are the averaged results of $10$ disks 
with different configuration of mass points, 
and the error bars are the standard deviations. 
As can be seen, $e$ decreases slightly with increasing colliding velocity and 
$e$ cannot reach $1$ as impact velocity decreases. 
We compare this result with the two-dimensional quasi-static theory 
of low speed impact\cite{brilliantov96,kuwabara,morgado}. 
In the two-dimensional quasi-static theory\cite{kuninaka_doctor}, 
the dynamical equation of motion for the deformation of 
the disk may be written as 
\begin{equation}\label{qst}
M \frac{d^2 \delta}{dt^2}=-P-\tau_{0} \frac{d P}
{d \delta} \frac{d \delta}{d t},
\end{equation}
where $M$ is the mass of the disk and $\tau_{0}$ is the time scale 
of dissipation.  
$P$ is obtained as a function of $\delta$ by 
numerically solving eq.(\ref{hertz_cnt}) for $P$. 
The last term in the right hand side is the dissipative force 
which is proportional to the velocity 
of the deformation\cite{brilliantov96,kuwabara}.
By introducing $\tau_{0}$ as a fitting parameter, 
we solve eqs.(\ref{hertz_cnt}) and (\ref{qst}) numerically 
with the initial conditions $\delta=0$ and 
$d\delta/dt=v$ to obtain the relation between $e$ and $v$. 
The solid line in Fig. \ref{quasi} (a) and (b) is 
the numerical result of eq.(\ref{qst}) with $\tau_{0}=0.011 R/c$. 
Figure \ref{quasi} (b) shows that our simulation data are well reproduced 
by the quasi-static theory in the low speed region ranging from $v/c=0.001$ to 
$v/c=0.007$. However, Fig. \ref{quasi} (a) shows that the quasi-static theory is 
no longer valid in the high speed region larger than $v/c=0.007$ in our simulation. 
The excitation of various internal modes originated from 
high speed impact causes such the discrepancy. 


\section{Concluding Remarks}

In summary, we reproduce Hertzian contact mechanism 
without introduction of any explicit dissipations. 
Our numerical results for low-speed impact are also consistent 
with the two-dimensional quasi-static theory. 
Through our investigation, 
we expect that we can obtain the further understanding on
the origin of the irreversibility from the reversible kinetic model. 

Our future tasks are (i) deriving the characteristic time $\tau_{0}$ 
in quasi-static impact from the microscopic mechanical model,  
(ii) extending our analysis to three dimensional problem, 
(iii) constructing a theory of impact for non quasi-static region, 
(iv) investigating the effect of the density distribution of colliding 
materials on their impact behavior, and (v) to check the effect of gravity 
for quasi-static impacts. For the last point, our preliminary results 
suggest that the restitution coefficient $e$ depends on the value of 
the gravitational acceleration.  

\section*{acknowledgments}
We would like to thank H. -D. Kim and N. Mitarai for their valuable comments. 
This study is partially supported by the Grant-in-Aid of 
Ministry of Education, Culture, Sports, Science and Technology (MEXT), 
Japan (Grant No. 15540393), the Grant-in-Aid for the 21st century COE 
"Center for Diversity and Universality in Physics" from MEXT, Japan, and 
the Grant-in-Aid from Japan Space Forum. 

\appendix
\section{Calculation of Elastic Moduli}\label{appA}
In this appendix, we show how to calculate elastic moduli, 
such as Young's modulus $E$ and Poisson's ratio $\nu$, of our model. 
In a two-dimensional isotropic medium, 
the energy density $E_{\rho}$ becomes 
\begin{equation}\label{comp}
E_{\rho} = 2 \epsilon^{2}(\lambda+\mu)
\end{equation}
in an isotropic compression while 
\begin{equation}\label{she}
E_{\rho} = 1/2 \mu \epsilon^{2}
\end{equation}
in a simple shear, 
where $\epsilon$ is the strain, and $\lambda$ and $\nu$ are 
Lam\'{e}'s constants. 

We calculate the energy densities of the disk in our model 
by adding an isotropic compression and a simple shear with 
the strain $\epsilon$. 
For an isotropic compression, 
by adding the small strain $\epsilon$ to each spring of the disk with 
the natural length $d_{j}$, 
we calculate the energy density of the disk as 
\begin{eqnarray}\label{eng_comp}
E_{\rho} &=& \sum_{j}\frac{1}{2A}\left(\frac{k_a}{2}(d_j \epsilon)^{2}+
\frac{k_b}{4}(d_j \epsilon)^{4}\right)\\\notag
&\simeq& \frac{k_a}{4A} \sum_{j} (d_{j} \epsilon)^{2},
\end{eqnarray}
where $A$ is the area of the disk. We neglect the fourth-order term. 
From eqs.(\ref{comp}) and (\ref{eng_comp}), 
on the assumption that our model is isotropic,  
we can calculate $\lambda+\mu$ as 
\begin{equation}\label{lm}
\lambda+\mu=\frac{k_{a}}{8 A} \sum_{i} d_{i}^{2}. 
\end{equation}

For a simple shear, 
by displacing the position of each mass point $(x_{i},y_{i})$ as 
$(x_{i}+\epsilon y_{i},y_{i})$ 
with the small strain $\epsilon$, 
we calculate the energy density of the disk as 
\begin{eqnarray}\label{eng_she}
E_{\rho} \simeq \frac{k_{a}}{4A} \sum_{j} \delta d_{j}^{2},
\end{eqnarray}
where $\delta d_{j}$ is the relative displacement from 
the natural length of the $j$-th spring. 
$\delta d_{j}$ is calculated as 
\begin{eqnarray}\label{dlt}
\delta d_{j}&=&\sqrt{\{x_{a}^{j}-x_{b}^{j} + \epsilon(y_{a}^{j}-y_{b}^{j})\}^{2}
+(y_{a}^{j}-y_{b}^{j})^{2}}-d_{j}\notag\\
&\simeq&\frac{\epsilon}{d_{j}}(x_{a}^{j}-x_{b}^{j})(y_{a}^{j}-y_{b}^{j}),
\end{eqnarray}
where $(x_{a}^{j},y_{a}^{j})$ and $(x_{b}^{j},y_{b}^{j})$ are 
the initial positions of both ends of the $j$-th spring. 
Thus, eq.(\ref{eng_she}) becomes 
\begin{equation}\label{eng_she2}
E_{\rho} \simeq \epsilon^{2} \frac{k_a}{4A} 
\sum_{j} \frac{1}{d_{j}^{2}}(x_{a}^{j}-x_{b}^{j})^{2}
(y_{a}^{j}-y_{b}^{j})^{2}. 
\end{equation}
From  eqs.(\ref{she}) and (\ref{eng_she2}), we can calculate $\mu$ as 
\begin{equation}\label{m}
\mu = \frac{k_a}{2 A} \sum_{j} \frac{1}{d_{j}^{2}}(x_{a}^{j}-x_{b}^{j})^{2}
(y_{a}^{j}-y_{b}^{j})^{2}. 
\end{equation}

Technically, we input the data 
for initial configuration of the connecting bonds of the disk 
into MATHEMATICA and put perturbations, 
such as the isotropic compression and the simple shear, 
to calculate the energy densities. 
Calculating $\lambda$ and $\mu$ from eqs.(\ref{lm}) and (\ref{m}) 
to substitute them into the two-dimensional relation,  
$E=4\mu(\lambda+\mu)/(2\mu+\lambda)$ and $\nu=\lambda/(2\mu+\lambda)$, 
we calculate Young's modulus $E$ and Poisson's ratio $\nu$ as 
$E = 0.773 k_{a}$ and $\nu = 0.336$, respectively.


\newpage
\end{document}